\documentclass[superscriptaddress,pre,floatfix,twocolumn,amsmath,amssymb,aps]{revtex4-1}

\usepackage{graphicx}
\usepackage{amsmath}
\usepackage{amssymb}
\usepackage{dcolumn}
\usepackage{color}
\usepackage{multirow}
\usepackage{bm}
\usepackage{subfigure}
\usepackage{tabularx}
\usepackage{booktabs}
\usepackage{appendix}
\usepackage{float}

\begin{document}
\title{Fermion sign problem and the structure of Lee-Yang zeros. \\II. Finite temperature results for a model system without interactions}
\author{Ran-Chen He}
\email{These authors contributed equally to this work.}
\affiliation{Interdisciplinary Institute of Light-Element Quantum Materials, Research Center for Light-Element Advanced Materials, and Collaborative Innovation Center of Quantum Matter, Peking University, Beijing 100871, P. R. China}
\affiliation{State Key Laboratory for Artificial Microstructure and Mesoscopic Physics, Frontier Science Center for Nano-optoelectronics and School of Physics, Peking University, Beijing 100871, P. R. China}
\author{Jia-Xi Zeng}
\email{These authors contributed equally to this work.}
\affiliation{State Key Laboratory for Artificial Microstructure and Mesoscopic Physics, Frontier Science Center for Nano-optoelectronics and School of Physics, Peking University, Beijing 100871, P. R. China}
\author{Shu Yang}
\affiliation{State Key Laboratory for Artificial Microstructure and Mesoscopic Physics, Frontier Science Center for Nano-optoelectronics and School of Physics, Peking University, Beijing 100871, P. R. China}
\author{Cong Wang}
\email{frankcongwang@pku.edu.cn}
\affiliation{State Key Laboratory for Artificial Microstructure and Mesoscopic Physics, Frontier Science Center for Nano-optoelectronics and School of Physics, Peking University, Beijing 100871, P. R. China}
\author{Qi-Jun Ye}
\email{qjye@pku.edu.cn}
\affiliation{Interdisciplinary Institute of Light-Element Quantum Materials, Research Center for Light-Element Advanced Materials, and Collaborative Innovation Center of Quantum Matter, Peking University, Beijing 100871, P. R. China}
\affiliation{State Key Laboratory for Artificial Microstructure and Mesoscopic Physics, Frontier Science Center for Nano-optoelectronics and School of Physics, Peking University, Beijing 100871, P. R. China}
\author{Xin-Zheng Li}
\email{xzli@pku.edu.cn}
\affiliation{Interdisciplinary Institute of Light-Element Quantum Materials, Research Center for Light-Element Advanced Materials, and Collaborative Innovation Center of Quantum Matter, Peking University, Beijing 100871, P. R. China}
\affiliation{State Key Laboratory for Artificial Microstructure and Mesoscopic Physics, Frontier Science Center for Nano-optoelectronics and School of Physics, Peking University, Beijing 100871, P. R. China}
\affiliation{Peking University Yangtze Delta Institute of Optoelectronics, Nantong, Jiangsu 226010, P. R. China}
\date{\today}

\begin{abstract}
Beyond the analysis of the Lee–Yang (LY) zero of $\xi$ at $0$~K presented by our previous work [He \textit{et. al.} Phys. Rev. E \textbf{113}, 24115 (2026)], it is important but intricate to understand how these zeros evolve with temperature ($T$). 
Here, we use an analytically solvable noninteracting one-dimensional particle-on-a-ring model to address this. 
We determine the trajectories of these zeros and analyze how their evolution with $T$ reshapes the analytic structure of the partition function. 
In particular, the zero originating from $\xi=-1$ at $T=0$ remains close to $-1$ at low $T$, where it governs the sign factor and strongly constrains continuation along the real $\xi$ axis.
This explains why both direct extrapolation and implicit schemes such as contour-based fitting can fail in the low-$T$ regime, even at high fitting order, while becoming reasonable again once the relevant zeros move away at higher $T$s. 
Furthermore, based on the polynomial structure of the partition function, we propose a new fitting strategy for low-$T$ fermionic properties.
The key is to first obtain reliable high-$T$ fermionic properties by continuing sign-problem-free data in $\xi\in[0,1]$ to $\xi=-1$, and then extend this information toward lower $T$ through $T$-fitting of the $\xi$-independent remainder $\phi(\beta)=Z_{\text{F}}$.
These results provide a solvable benchmark for diagnosing the validity of analytic continuation and suggest a possible route toward treating more realistic interacting fermionic systems.
\end{abstract}


\maketitle

\section{introduction}
In the previous article~\cite{He_2025}, we proposed an ensemble that exploits the exchange parity $\xi$ of indistinguishable particles as a parameter, and derived the partition function for a generalized many-body quantum system as a polynomial of $\xi$, as
\begin{equation}
    \label{anyonpoly}
    Z(N,\beta,\xi)=Z(\xi)=\sum_{j=0}^{N-1}c_j\cdot\xi^{j},
\end{equation}
where $\beta$ is the inversed temperature, $N$ is the number of particles, and $c_j$ represents for the associated coefficient.
Typically, $\xi$ is treated as a real variable, as it describes the exchange symmetry of indistinguishable bosons at $1$, indistinguishable fermions at $-1$, and distinguishable particles at $0$.
However, $\xi$ can also be extended to a complex variable, which describes the intermediate statistics of anyons in two-dimensional space at the other points on the unit circle $|\xi|=1$~\cite{leinaas1977,wilczek1982}, and Haldane's fractional exclusion statistics (FES) of ``excluson'' on the complex plane of $\xi$~\cite{haldane1991}.
Related one-dimensional anyon models and optical-lattice realizations provide further examples in which exchange statistics can be continuously tuned~\cite{batchelor2006,keilmann2011}.
The complexification of $\xi$ motivates us to study the distribution of the partition function zeros on this complex plane of $\xi$, which we call the Lee-Yang (LY) zeros of $\xi$~\cite{Yang_1952,Lee_1952,He_2025}. 
At $0$~K, we proved that for an arbitrary $N$-particle system, its zeros lie at 
\begin{align}
    \label{0zeros}
    &z_1=-1,\ z_2=-\frac{1}{2},\ z_3=-\frac{1}{3},\ \cdots,\nonumber\\
    &z_{N-1}=-\frac{1}{N-1},
\end{align}
independent of the specific forms of the inter-particle interactions and external potentials, as long as the quantum system has bound ground states and discrete energy levels in the low-energy region.
After that, by resorting to the third law of thermodynamics, we found that the LY zero of $\xi$ at $-1$ evolves according to
\begin{equation}
\label{z1}
z_1=-1+e^{-\beta E_0+o(\beta)},
\end{equation}
when the temperature ($T$) deviates from the $0$~K limit.
Here, $E_0=F_{0f}- F_{0b}$ is the free-energy difference between the fermionic and bosonic systems subjected to identical potential energies.
This nonzero difference implies a transition-like behavior when one connects the fermionic and bosonic systems at $0$~K along a chosen path on the real axis of $\xi$.
By ``transition-like'', we demonstrate that the fermionic and bosonic systems have different analytic forms of free energy.
Thus, direct analytical continuation between them, at least in the limit of $T$ approaching 0~K, is problematic. 
In realistic world, however, one lives at finite $T$s, where the thermodynamic behaviors can differ qualitatively from the 0~K limit in systems ranging from cold atoms~\cite{RevModPhys.80.885,Douglas_2015} to condensed matter~\cite{Bruus_2004,Coleman_2015,RevModPhys.73.33} and plasma or warm-dense-matter systems~\cite{thompson1964,birdsall1985,RevModPhys.84.1607,DORNHEIM20181,bonitz2024}.
As such, the universal 0~K distribution of the LY zeros of $\xi$ at $-1$, $-1/2$, $\cdots$, $1/(N-1)$ must be generally distorted.
Inter-particle interactions and external potentials will also reshape this finite-$T$ distribution of the LY zeros of $\xi$ in a system-dependent way.
This breakdown of the 0~K universality, together with the rich behaviors intrinsic to finite-$T$ thermodynamic systems, is exactly the motivation of this work.
To isolate finite-$T$ effects from interaction effects and provide a clear baseline for understanding how the zeros evolve with $T$, we choose the simplest model---a non-interacting indistinguishable one-dimensional~(1D) particle-on-a-ring model, where analytic expressions for the partition function and energies are available.
Based on the evolution of these zeros, we investigate how the distribution of the LY zeros of $\xi$ at finite $T$s governs the validity and limitations of the analytic continuation scheme, as used in earlier state-of-the-art computer simulations of the fermion sign problem (FSP)~\cite{RN511,RN506,Xiong2022,Xiong2023,xiong2024,Dornheim2023,dornheim2024,dornheim2024a,dornheim2025,dornheim2025a,dornheim2025c,dornheim2026,Tommaso2025,morresi2025a,xiong2025b,fan2025a}.
In the subsequent paper, we will extend the analysis to interacting systems by numerical methods, with particular focus on how interactions modify the finite-$T$ behavior of LY zeros of $\xi$.
We hope this step-by-step strategy---from the universal 0~K patterns, via the exactly solvable non-interacting finite-$T$ benchmarks, to realistic interacting systems---builds a comprehensive framework for understanding FSP and offers useful insights for simulations of realistic fermion systems.
This paper proceeds as follows.
We first introduce the model in Sec.~\ref{1Dmod} and use it to track the finite-$T$ trajectories of the LY zeros.
Sec.~\ref{FPat-1} then connects these trajectories to thermodynamic behavior and to the reliability of fermionic-property extrapolation.
In Sec.~\ref{CEC}, we examine constant-energy contours and show that, although they are not formulated as a direct continuation in $\xi$, they still amount to an implicit extrapolation scheme whose regime of validity is controlled by the same LY-zero structure.
Sec.~\ref{pfstructure} turns to the partition function itself and analyzes its structure near the fermionic point.
Finally, Sec.~\ref{attempt} discusses possible routes for obtaining low-$T$ fermionic properties, before Sec.~\ref{Conclusion} summarizes the main conclusions.

\section{Lee-Yang zeros of $\xi$ at finite Temperatures}
\label{1Dmod}
We put $N$ indistinguishable noninteracting particles on a 1D ring.
In this setup, the single-particle energy eigenvalues are
\begin{equation}
    E_n=\frac{n^2h^2}{2mL^2},
\end{equation}
where $h$ is Planck's constant, $m$ is the particle mass, and $L$ is the length of ring.
The many-body eigenstates can then be constructed from these single-particle states with exchange symmetry enforced, e.g., as Slater determinants for fermions and permanents for bosons.
For the simplest case, $N=1$, the partition function of the system reads
\begin{equation}
    \label{EZ}
    Z(N=1,\beta,\xi)=\sum_{n=-\infty}^\infty e^{-\frac{\beta n^2h^2}{2mL^2}}.
\end{equation}
Since there is no exchange in a one-particle system, $\xi$-dependence does not appear on the right-hand-side of this equation.
In the path-integral formulation, the partition function can also be written as
\begin{equation}
    \label{Z11}
        Z(N=1,\beta,\xi) =\int dR_1 \langle R_1 |e^{-\beta \hat H^{(1)}}|R_1\rangle,
\end{equation} 
where $\hat H^{(1)}$ is single-particle Hamiltonian.

When $N>1$, exchange symmetry becomes explicit.
In the path-integral representation, this means that the statistical weight is no longer built only from ring-polymer topologies of labelled-particle, but must also include contributions from different permutation cycles~\cite{Hirshberg2019,Hirshberg2020,Feldman2023,higer2025}.
In Ref.~[\onlinecite{He_2025}], the expansion of the partition function as a polynomial of $\xi$ was explained in detail.
Here, to avoid being repetitive, we just summarize the central equations.
The key point is that the partition function of the $N$-particle system can be decomposed into contributions from different $Z_Q$s by
\begin{equation}
   \label{ZNPartition}
     Z^{(N)}(\beta,\xi)=\sum_{Q} \xi^{\sigma(Q)} \frac{F_Q}{N!} Z_Q,
\end{equation}
where
\begin{equation}
   \label{ZQ}
     Z_Q=\int d\mathbf{R} \langle p\mathbf{R} |e^{-\beta \hat H}|\mathbf{R}\rangle,
\end{equation}
and 
\begin{equation}
\label{FQN}
F_Q=\frac{N!}{\big(1^{\gamma_1}\gamma_1!\big)\big(2^{\gamma_2}\gamma_2!\big)\cdots\big(N^{\gamma_N}\gamma_N!\big)}.
\end{equation}
Here, $Q$ represents one class of the permutation group, $F_Q$ is the number of elements in this class, and $p$ denotes one element in it.
$\sigma(Q)$ means the minimal number of pair permutations to restore the identity element from the elements in $Q$.
One property of the permutation group is that each class of it can be denoted by the cycling structure of the corresponding permutation, i.e. $1^{\gamma_1}2^{\gamma_2}\cdots N^{\gamma_N}$, with $\gamma_m$ denoting the number of $m$-cycles (permuting all the $m$ elements in a single cycle) and satisfying
\begin{equation}
   \label{cycle1}
     \sum_{m=1}^{N} m\cdot \gamma_m = N.
\end{equation}
Therefore, the label of $Q$ in Eqs.~[\ref{ZNPartition}] to [\ref{FQN}] can be replaced by its cycling structure $1^{\gamma_1}2^{\gamma_2}\cdots N^{\gamma_N}$.
For example, when $1^{\gamma_1}2^{\gamma_2}\cdots N^{\gamma_N}=1^{0}2^{0}\cdots N^{1}$ (abbreviated as $N^{1}$ by omitting zero terms), the corresponding $Z_Q$ is
\begin{equation}
   \label{ZN1}
     Z_{N^1}=\int d\mathbf{R} \langle R_1 R_2 \cdots R_N |e^{-\beta \hat H}|R_2\cdots R_N R_1\rangle,
\end{equation}
where $\mathbf{R} = \{R_1,\cdots,R_N\}$ is the set of the coordinates of the particles, and thus $d\mathbf{R}=\prod_{j=1}^{N}dR_j$.
Here, Eq.~\eqref{ZN1} just corresponds to one topology of the polymer where the $N$ identical particles are connected into one single largest ring.
In Sec. IID of Ref.~[\onlinecite{He_2025}], we have explained that for a noninteracting system, the contribution from each permutation class to 
the partition function can be factorized as
\begin{equation}\label{ZN11b}
Z_{1^{\gamma_1}2^{\gamma_2}\cdots N^{\gamma_N}}=\prod_{j=1}^{N}(Z_{j^1})^{\gamma_j}.
\end{equation}
The key task here, therefore, is to derive the analytic form of $Z_{j^1}$ using Eq.~(\ref{ZN1}).
In Eq.~[\ref{ZNPartition}], exchange symmetry is already encoded in the factor of $\xi^{\sigma(1^{\gamma_1}2^{\gamma_2}\cdots N^{\gamma_N})}$.
Therefore, the bra $\langle R_1 R_2\cdots R_N |$ and ket $|R_2\cdots R_N R_1\rangle$, as in Eq.~\eqref{ZN1}, can be understood as eigenstates 
of a many-body system of ``labelled'' particles~\cite{He_2025}.
Accordingly, the many-body coordinate basis can be factorized into
\begin{equation}
    |R_1 R_2 \cdots R_N\rangle=|R_1\rangle|R_2\rangle \cdots|R_N\rangle.
\end{equation}
This property is crucial for the following simplifications and discussion.
Since there is no interaction between particles in our model, the Hamiltonian is a sum of identical single-particle terms and we can rewrite Eq.~\eqref{ZN1} into 
\begin{equation}
\begin{aligned}
    \label{ZN11}
       Z_{N^1} & =\int d\mathbf{R} \langle  R_1|e^{-\beta \hat H^{(1)}}| R_2\rangle\langle R_2|e^{-\beta \hat H^{(2)}}| R_3\rangle \\ 
&\quad  \cdots\langle R_N|e^{-\beta \hat H^{(N)}}| R_1\rangle\\
&= \int dR_1\langle R_1 |e^{-N \beta \hat H^{(1)}}|R_1\rangle=Z_{1^1}(N\beta).
\end{aligned}
\end{equation}
$Z_{N^1}$ thus equals $Z_{1^1}(N\beta)$, which represents a single particle on one ring at the effective temperature $T/N$. 
\begin{figure}[b]
    \centering
    \includegraphics[width=0.85\linewidth]{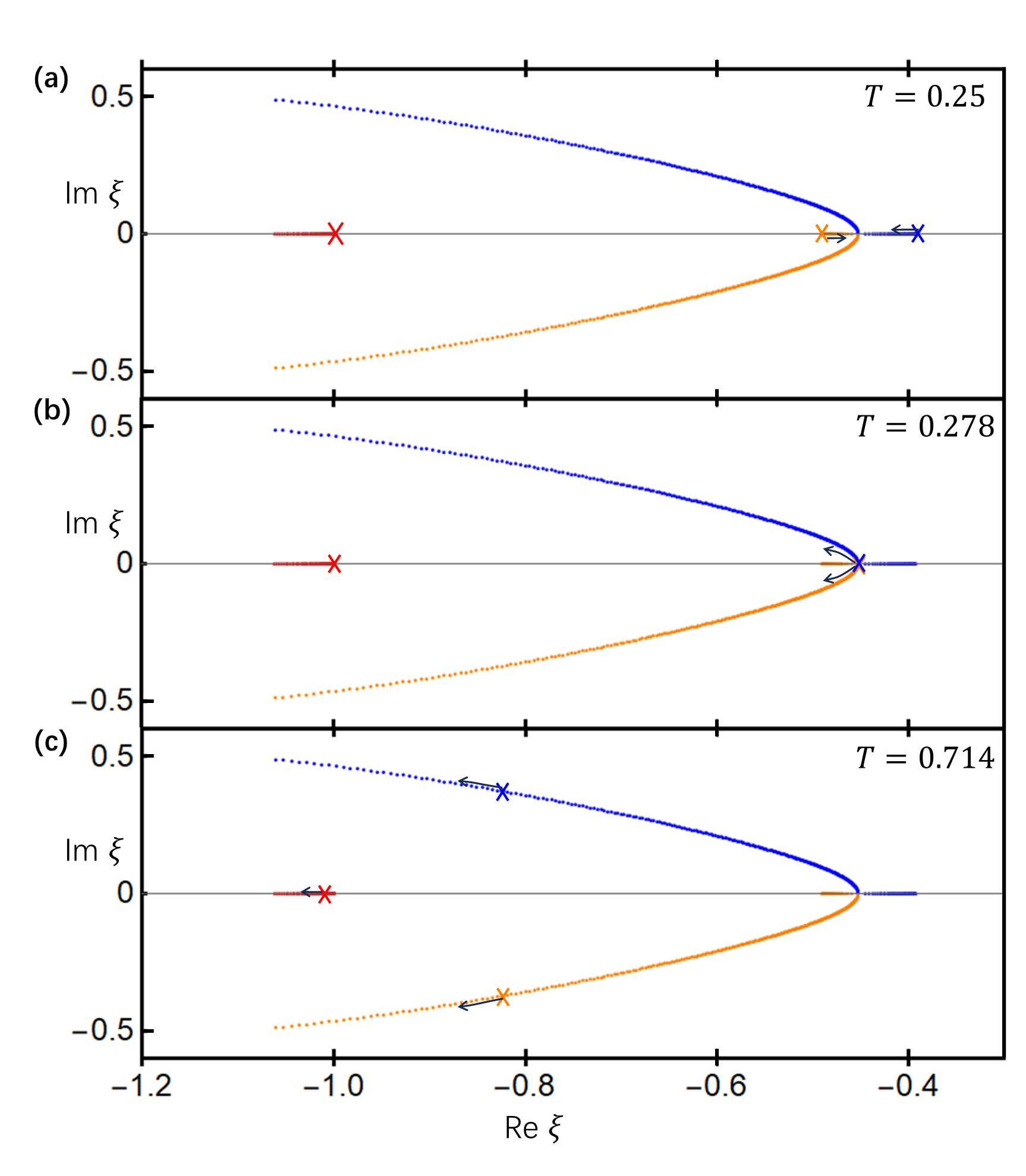}
    \caption{The LY zeros of $\xi$ at different $T$s for the noninteracting $N=4$ system on a 1D ring.
    For $N=4$, there are three LY zeros of $\xi$ and thus three trajectories (dotted lines connecting the same zero at different $T$s), shown in the background.
    We identify the zeros by continuously tracking their $T$-dependent evolution, and label them in red, orange, and blue colors.
    The temperature step is $\Delta T=0.01$ from $T=0.25$ to $T=1$, in units of $h^2/(2mk_BL^2)$.
    The zeros corresponding to (a) $T=0.25$, (b) $T=0.278$, and (c) $T=0.714$ are highlighted as larger crosses, with black arrows indicating the directions of motion as $T$ increases.
    For low-$T$ cases, the zero originating from $-1$ remains very close to $-1$ according to Eq.~\eqref{z1}; therefore, its arrow is omitted.}
    \label{zeros}
\end{figure}

\begin{figure}[htbp]
    \centering
    \includegraphics[width=0.85\linewidth]{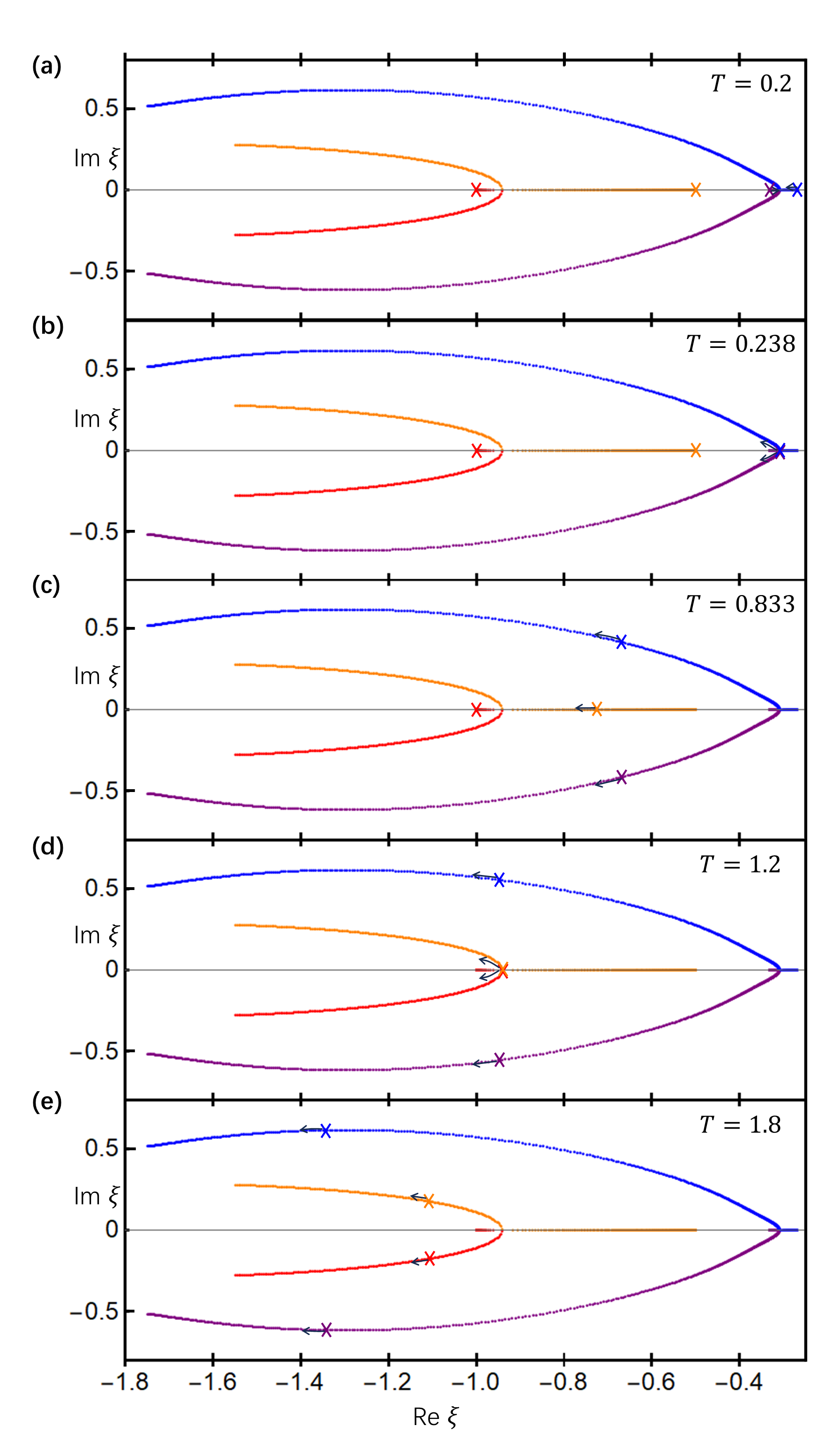}
    \caption{The LY zeros of $\xi$ at different $T$s for the noninteracting $N=5$ system on a 1D ring. 
    The plotting conventions are the same as in Fig.~\ref{zeros}. 
    Here, four zeros are shown (red, orange, blue, and purple), with $\Delta T=0.01$ from $T=0.2$ to $T=3$.
    The zeros at (a) $T=0.2$, (b) $T=0.252$, (c) $T=1$, (d) $T=1.65$, and (e) $T=1.8$ are highlighted as larger crosses.
    Similar to Fig.~\ref{zeros}, arrows indicate motion with increasing $T$, except for the low-$T$ zero originating from $-1$, whose arrow is omitted due to Eq.~\eqref{z1}.}
    \label{zeros5}
\end{figure}

Combining Eqs.~\eqref{ZN11b} and~\eqref{ZN11}, we obtain explicit analytic expressions for the partition functions of the $N=1$ to $5$ systems:
\begin{equation}
    \label{nointer123}
    \begin{split}
    Z^{(1)}(\beta,\xi)&=Z_{1^1}(\beta),\\
    Z^{(2)}(\beta,\xi)&=\frac{1}{2}\big[Z_{1^1}^2(\beta)+\xi Z_{1^1}(2\beta)\big],\\
    Z^{(3)}(\beta,\xi)&=\frac{1}{6}\big[Z_{1^1}^3(\beta)+3\xi Z_{1^1}(\beta)Z_{1^1}(2\beta)+2\xi^2Z_{1^1}(3\beta)\big],\\
    Z^{(4)}(\beta,\xi)&=\frac{1}{24}\big[Z_{1^1}^4(\beta)+6\xi Z_{1^1}^2(\beta)Z_{1^1}(2\beta)+\\
&\quad 8\xi^2Z_{1^1}(\beta)Z_{1^1}(3\beta)+3\xi^2Z_{1^1}^2(2\beta)+\\
&\quad 6\xi^3Z_{1^1}(4\beta)\big],\\
    Z^{(5)}(\beta,\xi)&=\frac{1}{120}\big[ Z_{1^1}^5(\beta)+10\xi Z_{1^1}^3(\beta)Z_{1^1}(2\beta)+\\
&\quad 20\xi^2Z_{1^1}^2(\beta)Z_{1^1}(3\beta)+15\xi^2 Z_{1^1}(\beta)Z_{1^1}(2\beta)+\\
&\quad 30\xi^3Z_{1^1}(\beta)Z_{1^1}(4\beta)+20\xi^3 Z_{1^1}(2\beta)Z_{1^1}(3\beta)+\\
&\quad 24\xi^4 Z_{1^1}(5\beta)\big].
    \end{split}
\end{equation}
Finally, substituting Eq.~\eqref{EZ} into Eq.~\eqref{nointer123} yields the finite-$T$ partition functions in a fully analytical form.
Using these analytic partition functions, we track the trajectories of LY zeros of $\xi$ in the complex plane for $N=4$ and $5$.
These trajectories are constructed by calculating the LY zeros of $\xi$ at equal $T$ intervals and connecting the positions of the same zero at different $T$s, 
as shown by colored dotted lines in Figs.~\ref{zeros} and \ref{zeros5}.
At moderate $T$s, all zeros originating from $\xi = -1, -1/2, -1/3$ (and $-1/4$ for $N=5$) stay on the real axis between $-1$ and $0$, as shown in Figs.~\ref{zeros}(a)-(b) and~\ref{zeros5}(a)-(b).
As $T$ increases, the zeros move away from the real axis, as shown by the panel corresponding to $T=0.714h^2/(2mk_BL^2)$ for $N=4$ (Fig.~\ref{zeros}(c)) 
and $T=1.8h^2/(2mk_BL^2)$ for $N=5$ (Fig.~\ref{zeros5}(e)).
The zero originating from $\xi=-1$ is the last one to leave the real axis. 
It is worth noting that a parity constraint also exists for the early-stage evolution of $z_1$ when $T$ departs from 0~K.
This is due to the fact that the partition function must be positive for both fermionic and bosonic systems ($\xi=1$ and $\xi=-1$). 
As a consequence, along the real axis from $-1$ to $1$, the number of zero crossings must be even, with a double root counted twice. 
Thus, for even $N$ (e.g., $N=4$), the total number of zeros is odd, and $z_1$ is pushed to the left of $-1$ before complex pairing occurs (Fig.~\ref{zeros}(c)).
For odd $N$ (e.g., $N=5$), by contrast, $z_1$ can remain on the real axis until it meets another zero, after which they leave the real axis as a conjugate pair (Figs.~\ref{zeros5}(d)-(e)). 
This difference also explains the distinct constant-energy-contour behaviors, which will be discussed later in Sec.~\ref{CEC}.

\begin{figure}[htbp]
    \centering
    \includegraphics[width=0.85\linewidth]{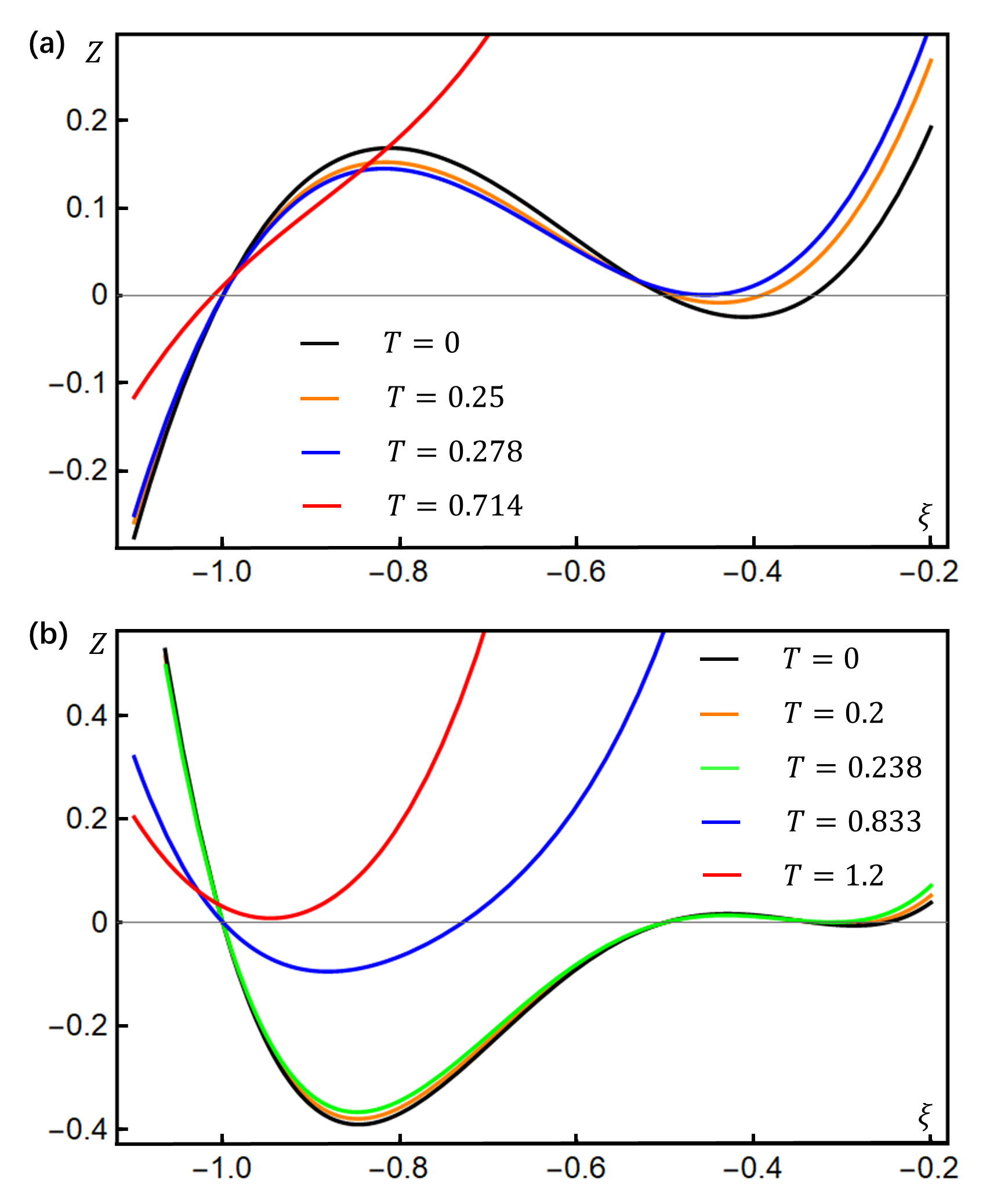}
    \caption{(a) Curves of $Z(\xi)$ on the real $\xi$ axis for $N=4$. LY zeros of $\xi$ are exactly the intersections with $Z=0$. At $T=0$, there are three roots, as $\xi =-1,-1/2, -1/3$. As $T$ increases, two neighboring real roots merge into a double root at $\xi\approx-0.45$ (around $T=0.278$), and then leave the real axis as a complex-conjugate pair, so that only one real zero remains. (b) Curves of $Z(\xi)$ for $N=5$. Consistent with the parity constraint discussed in the text, the four roots initialized at $\xi=-1,-1/2,-1/3,-1/4$ leave the real axis in two pairs; consequently, no real zero remains at sufficiently high $T$ (here $T\gtrsim1.2$).}
    \label{poly_curve}
\end{figure}

This mechanism can be interpreted graphically in Fig.~\ref{poly_curve}.
Viewing $Z(\xi)$ as a polynomial curve on the real $\xi$ axis, LY zeros are exactly the intersections with the $x$ axis corresponding to $Z=0$.
As $T$ varies, the polynomial coefficients change, deforming the curve of $Z(\xi)$.
When a segment of the curve is lifted upward, two neighboring intersections (two zeros) approach each other and gradually merge at a tangency point (as a double root).
When the intersections finally disappear from the real axis, the corresponding zeros must appear as a complex-conjugate pair, as shown in Figs.~\ref{zeros} and ~\ref{zeros5}.

\section{Thermodynamic properties versus $\xi$ and the validity of extrapolation methods}
\label{FPat-1}

In Ref.~[\onlinecite{He_2025}], we analyzed the distribution of LY zeros of $\xi$ at 0~K and argued that it may obstruct the analyticity of $\xi$-dependent thermodynamic quantities at low $T$s.
Our logic is simple and follows the LY-zero perspective, which has also been used to analyze supercritical phenomena and realistic molecular phase-transition problems~\cite{Yang_1952,Lee_1952,Ouyang_2023,liu2025}: when LY zeros of $\xi$ lie on the real axis, the observables become nonanalytic at these points and the extrapolation 
scheme should break down; when the zeros are close to the real axis, the extrapolation should also be strongly affected, analogous to the supercritical behavior 
near a critical point~\cite{Fisher1965,Ouyang_2023}.
Since the zeros stay on the real axis at low $T$s and deviate from it as $T$ increases (Sec.~\ref{1Dmod}), the validity of the extrapolation methods should be $T$-dependent 
and must be tested carefully.
Using the finite-$T$ trajectories of the LY zeros of $\xi$ and the analytical form of the thermodynamic quantities given in Sec.~\ref{1Dmod}, we can now investigate in a quantitative manner how the thermodynamic observables and the extrapolation methods for the FSP~\cite{Xiong2022, Xiong2023, Dornheim2023, Tommaso2025, dornheim2026} are affected by these zeros. 

\begin{figure}[htbpb]
    \centering
    \includegraphics[width=0.95\linewidth]{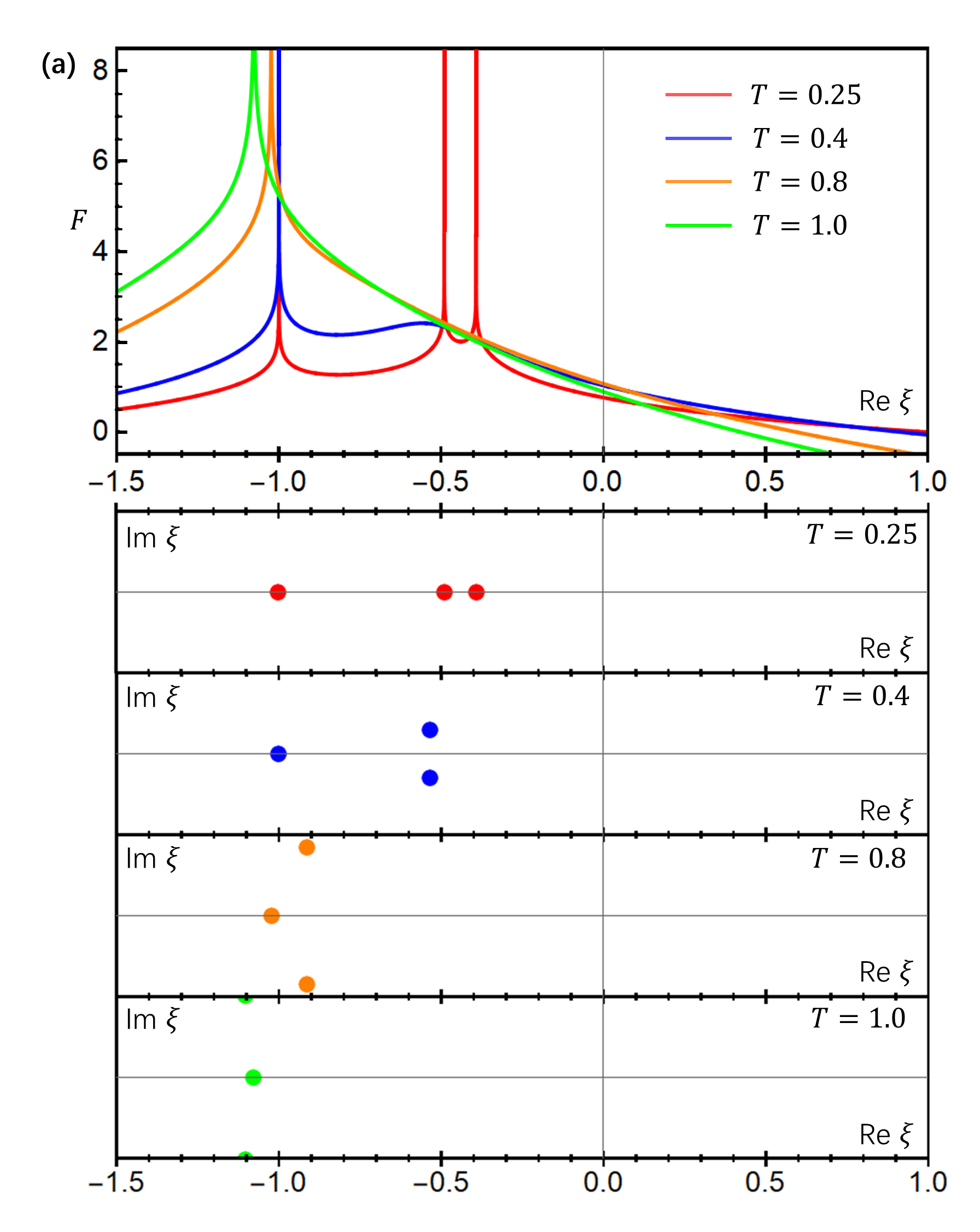}
	\includegraphics[width=0.95\linewidth]{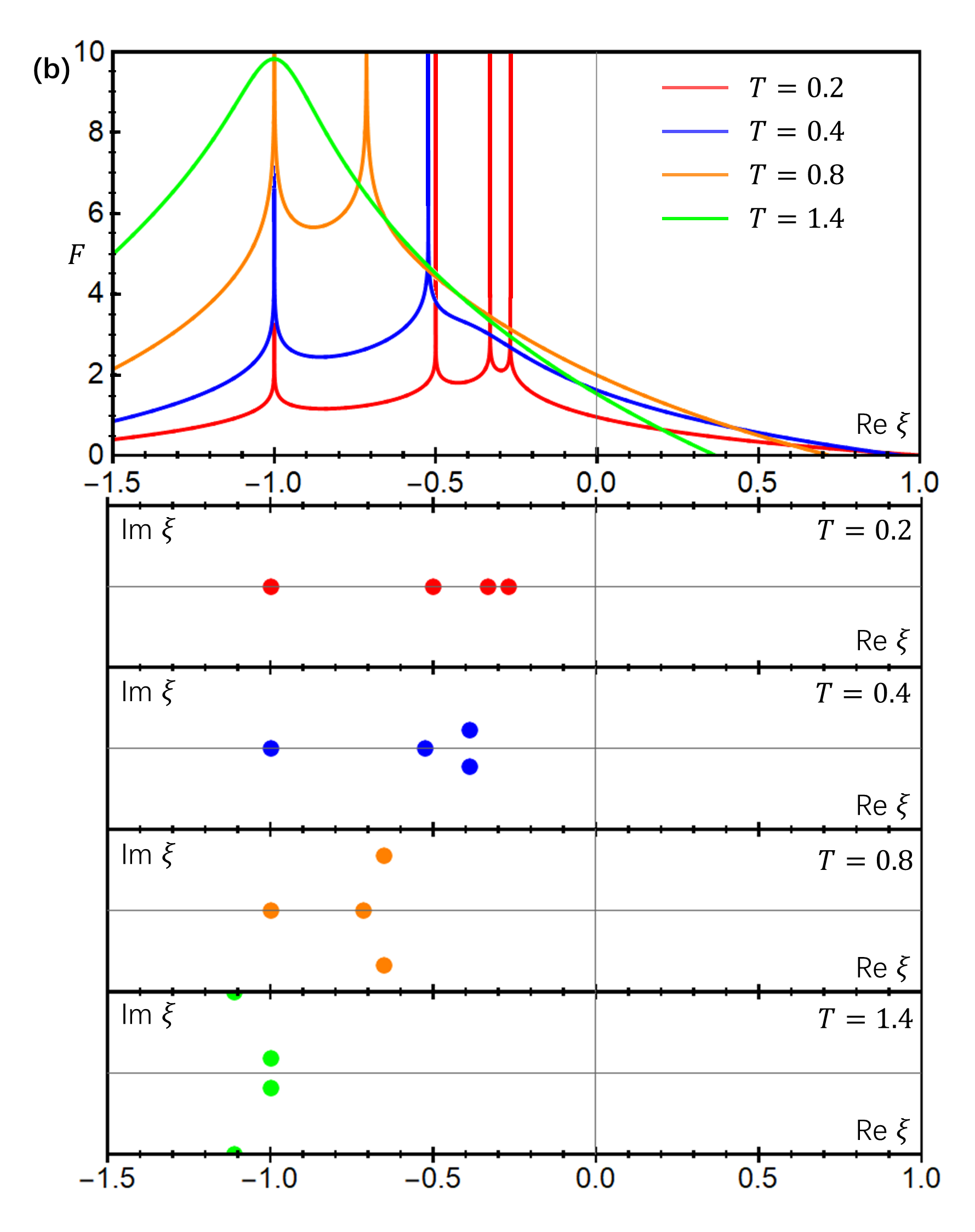}
    \caption{Free energy $F(T,\xi)$ versus $\xi$ for (a) $N=4$ and (b) $N=5$, with LY zeros depicted at the corresponding $T$s (different colors). In accordance with the main text, zeros on the real axis correspond to nonanalytic points (sharp divergences), while zeros close to the real axis still distort the free energy profile. The unit of free energy is $h^2/(2mL^2)$.}
    \label{Fxi}
\end{figure}

To this end, we focus on two key thermodynamic quantities of the fermionic system: the free energy $F(T,\xi=-1)$ and the sign factor $\langle s \rangle_{\text{B}}$.
The free energy is proportional to the logarithm of the partition function.
By recasting $Z(\beta,\xi)$ in Eq.~\eqref{anyonpoly} into a product of terms related to zeros $\{z_i\}$, as
\begin{equation}
\label{pfz0}
Z(\beta,\xi)=\frac{Z_{N^1}}{N}\prod_{i=1}^{N-1}(\xi-z_i),
\end{equation}
where $z_i$ labels the zeros sequentially by their 0~K positions,
we formulate the free energy as
\begin{equation}
\begin{aligned}
    \label{energy1}
    F(T,\xi)&=-k_{\text{B}}T\ln Z(T,\xi)\\
&=-k_{\text{B}}T\left(\ln Z_{N^1}+\sum_{i=1}^{N-1}\ln |z_i-\xi| -\ln N\right).
\end{aligned}
\end{equation}
In Fig.~\ref{Fxi}, we plot the free energy as a function of real $\xi$, as well as the corresponding LY zeros of $\xi$ on the complex plane of it.
This quantitatively confirms the expectation of our previous paper~\cite{He_2025}.
In Eq.~\eqref{energy1}, the term $\ln |z_i-\xi|$ admits an electrostatic analogy~\cite{Lee_1952}: each $z_i$ affects the free energy $F(T,\xi)$ like an infinitely long charged line contributing to the electrostatic potential at a point on the complex plane of $\xi$.
When any $z_i$ stays on the real axis such that a real $\xi$ can equal it, $|z_i-\xi|$ equals $0$ and $F(T,\xi)$ diverges at this point.
This leads to sharp peaks in red in Figs.~\ref{Fxi}(a) and (b).
When $z_i$ leaves the real axis, this electrostatic effect is determined by the finite value of $|z_i-\xi|$: the sharp divergence gradually turns into a smooth turning point (e.g., the blue curve at $T=0.4$ around $\xi=-0.5$ in Fig.~\ref{Fxi}(a)) and eventually becomes subtle (e.g., the green curve at $T=1.4$ behaves smoothly for $\xi>-1$ in Fig.~\ref{Fxi}(b)).
Accordingly, when the zeros are sufficiently far from the real axis, continuation of the free energy along the real axis of $\xi$ from $(0,1)$ to $-1$ can be trusted, and $F(T,\xi)$ can be treated as an analytic function of $\xi$ between $-1$ and $1$.
\begin{figure}[b]
    \centering
    \includegraphics[width=0.9\linewidth]{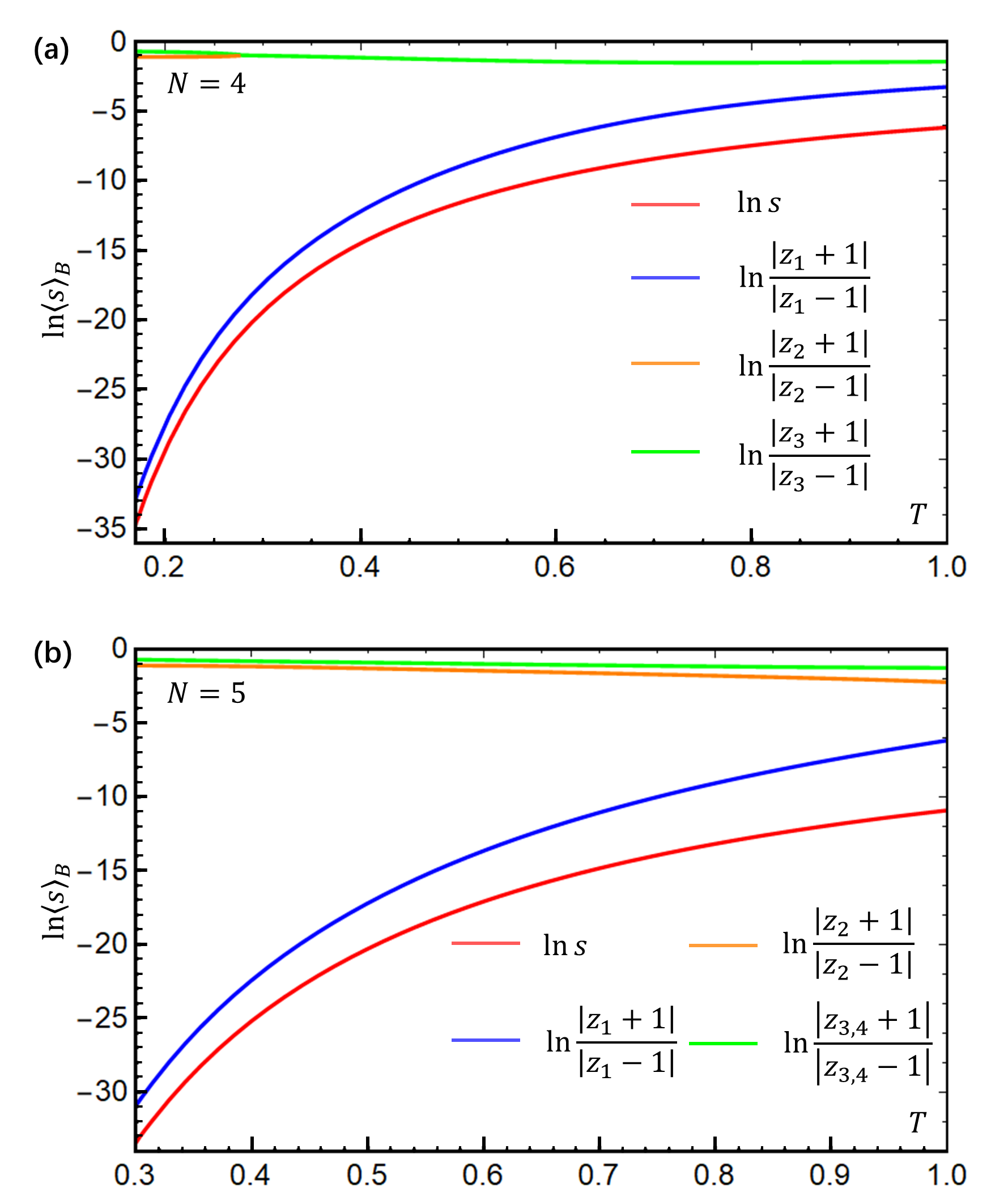}
    \caption{Sign factor $\langle s \rangle_B$ and LY zero $z_1$ for $N=4$ and $5$ on a logarithmic scale. 
    The blue curves show the contribution from $z_1$ as temperature increases, and they exhibit behavior very similar to the red curves, which represent the sign factor $\langle s \rangle_B$ on the same logarithmic scale. The temperature unit is $h^2/(2mk_BL^2)$.}
    \label{zeross2}
\end{figure}
In addition to the free energy, the sign factor $\langle s \rangle_{\text{B}}$ can also be expressed and analyzed within a $\xi$-dependent formulation.
Substituting Eq.~\eqref{pfz0} into $\langle s \rangle_{\text{B}}=Z_{\text{F}}/Z_{\text{B}}$ yields
\begin{equation}
    \label{signfactor}
    \langle s \rangle_{\text{B}}=\frac{Z_{\text{F}}}{Z_{\text{B}}}= \frac{Z(\beta,-1)}{Z(\beta,1)}=\prod_{i=1}^{N-1}\frac{z_i+1}{z_i-1}.
\end{equation}
$\langle s \rangle_{\text{B}}$ is affected by all zeros according to their distance to $\xi=-1$, and is thus dominated by $z_1$ (the zero closest to $\xi=-1$).
To show this pictorially, in Fig.~\ref{zeross2} we plot $\ln \langle s \rangle_{\text{B}}$ as a function of $T$ for our model systems and decompose its contribution from different $z_i$.
As $T$ decreases, $\ln \langle s \rangle_{\text{B}}$ becomes increasingly small, which is exactly the source of the numerical difficulty in sampling the sign factor.
In such cases, the blue curve dominates the red one in both panels of Fig.~\ref{zeross2}.
A small sign factor (close to zero) means that reweighting bosonic statistics becomes of limited use, or even futile in the 0~K limit, for extracting fermionic properties.
This provides an alternative explanation for the well-known convergence problem of the reweighting method.
Only when $z_1$ moves away from $-1$, or even leaves the real axis at high $T$s, does $\langle s \rangle_B$ become appreciable and the reweighting scheme become effective.
The discussion in the last paragraph is clear as it stands. 
However, one special case requires further explanation. 
Due to the parity constraint discussed in Sec.~\ref{1Dmod}, there is a special even-$N$ scenario in which $z_2$ moves close to $z_1$ before the pair leaves the real axis.
When $|z_2+1|$ becomes comparable to $|z_1+1|$, the leading contribution in Eqs.~\eqref{energy1} and~\eqref{signfactor} is no longer controlled by $z_1$ alone, but by the combined $z_1$ and $z_2$ terms.
This leads to a qualitative change in the behavior, although the overall low-$T$ trend and the conclusions of this section remain unchanged.
\section{The constant-energy contour}
\label{CEC}
\begin{figure}[htbp]
    \centering
    \includegraphics[width=0.9\linewidth]{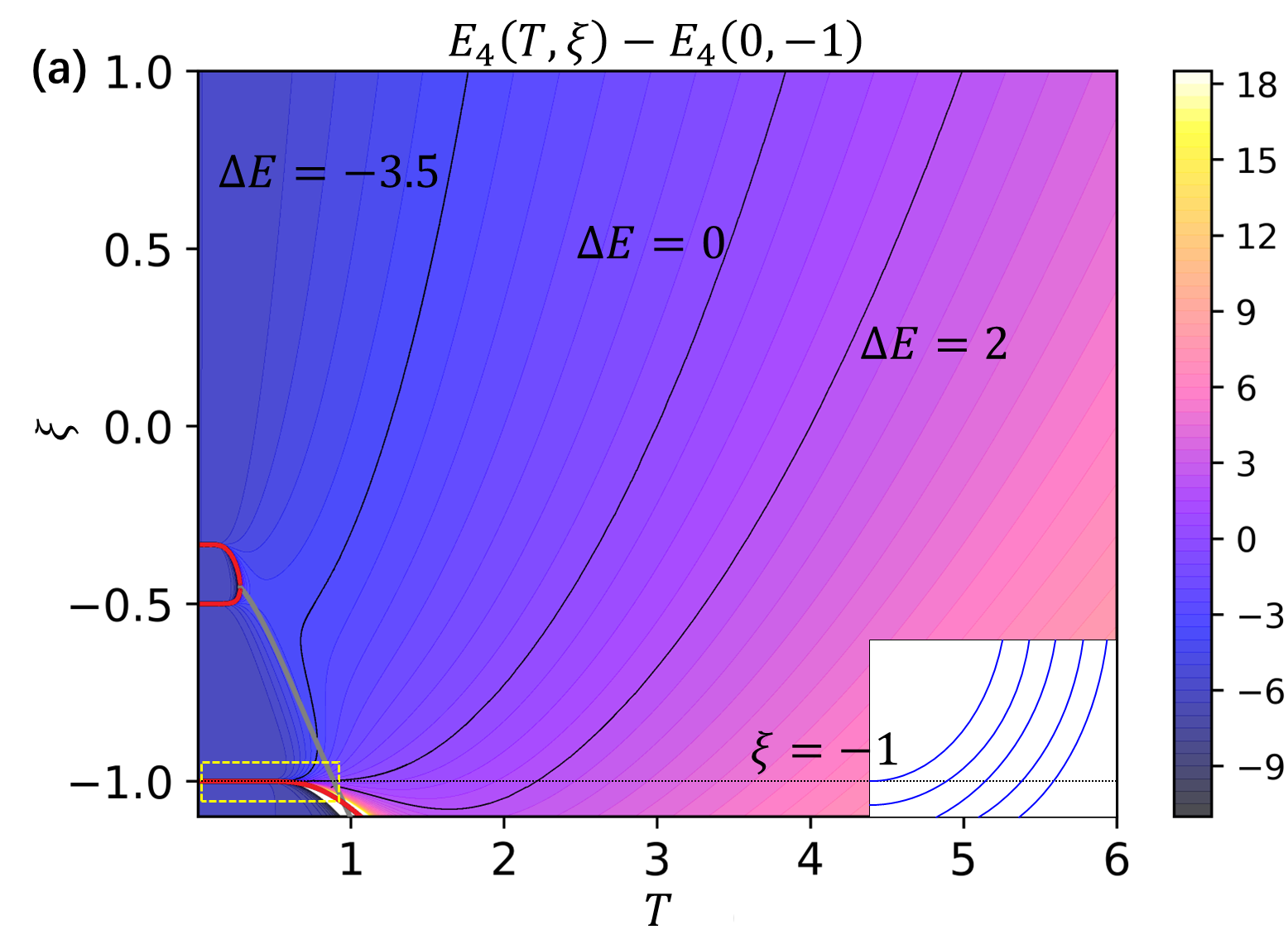}
\includegraphics[width=0.9\linewidth]{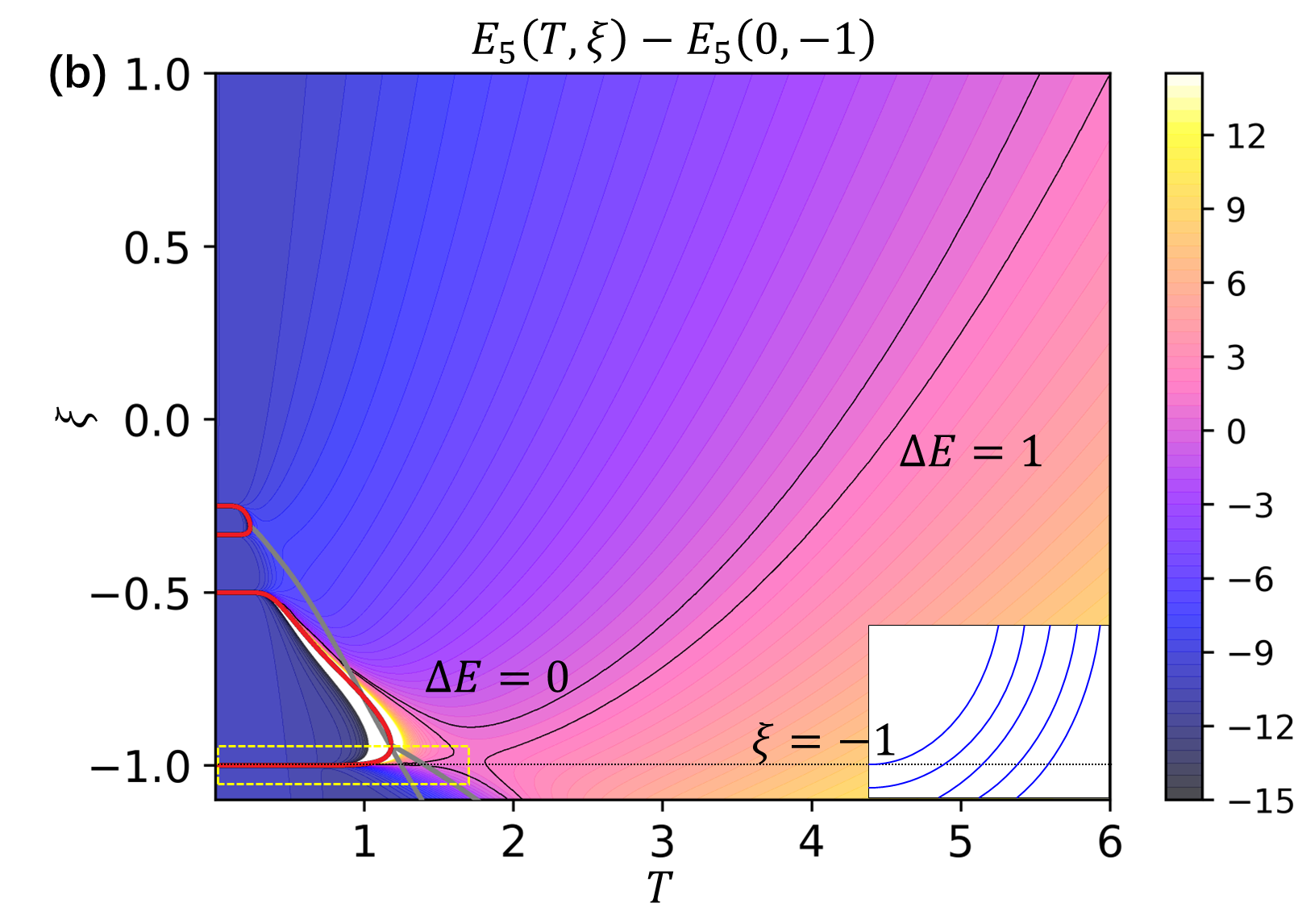}
    \caption{Constant-energy contours in the $\xi$-$T$ plane for $N=4$ and $N=5$. Black lines denote representative contours of fixed $\Delta E$. Thick red curves show trajectories of LY zeros on the real axis, and thick gray curves show the trajectories of the real part of $\xi$ after the zeros move off the real axis. Contour distortion near the red curves highlights the low-$T$ breakdown of extrapolation discussed in the text. The unit of $\Delta E$ is $h^2/(2mL^2)$.}
    \label{zeross}
\end{figure}

\begin{figure}[htbp]
    \centering
    \includegraphics[width=0.95\linewidth]{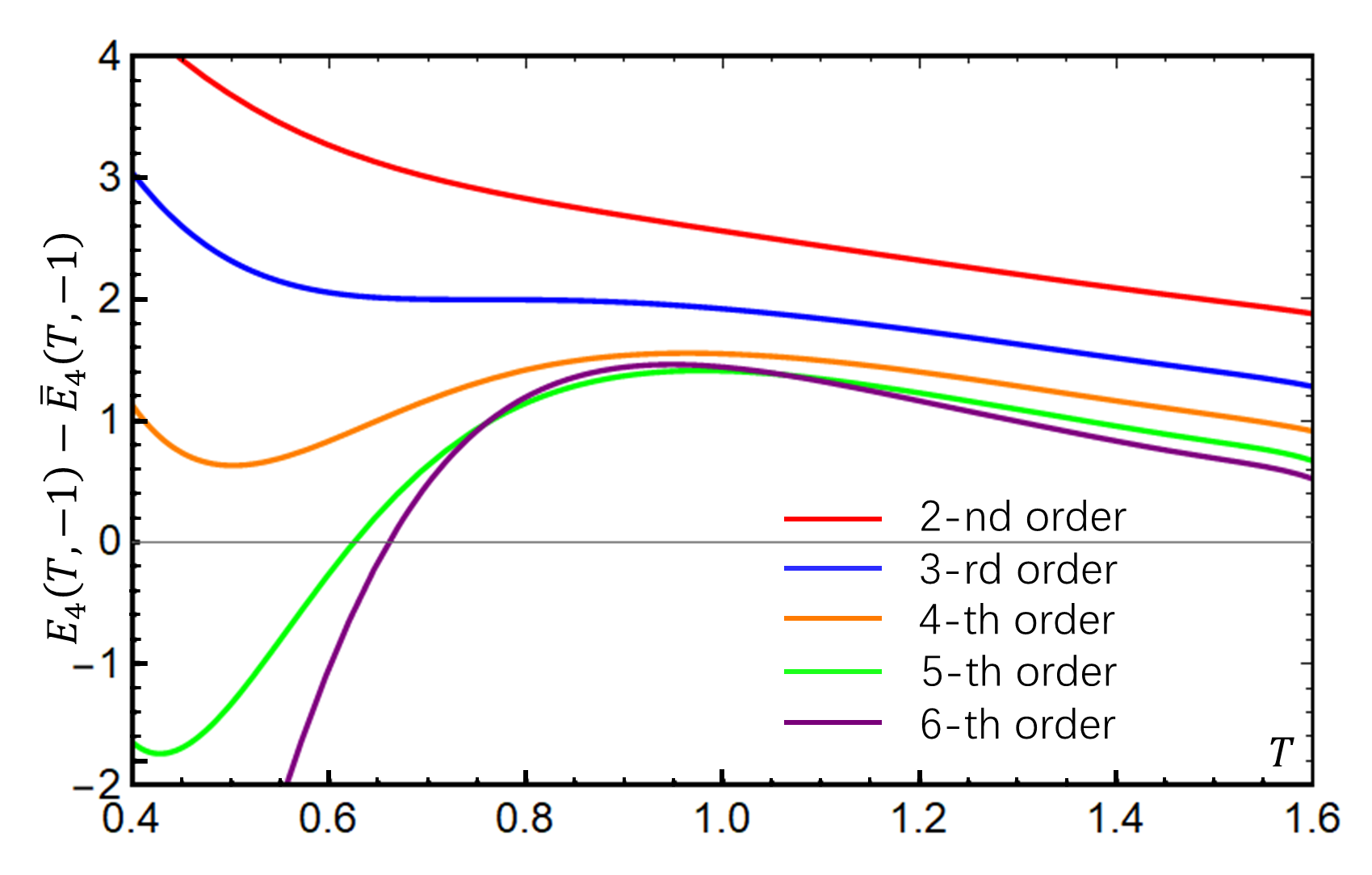}
    \caption{Difference between the fitted energy from the formula of Xiong \textit{et al.} [Eq.~\eqref{Xiong}] and the exact energy at $\xi=-1$ for $N=4$, shown for different truncation orders of the expansion. The unit of energy difference is $h^2/(2mL^2)$.}
    \label{fit}
\end{figure}

Building on the previous section, where we established the regime in which analytical extrapolation is valid, we now turn to the method proposed by Xiong \textit{et al.}~[\onlinecite{Xiong2023}] for estimating finite-$T$ fermionic energies from constant-energy contours.
Its fitting form is
\begin{equation}
\label{Xiong}
\xi_E(T)=a_0(E)+a_2(E)T^2+\sum_{i>2}a_i(E)T^i.
\end{equation}
For each fixed energy $E$, this scheme fits $\xi$ as an analytic function of $T$.
Although it is not a direct extrapolation along the real axis of $\xi$, it is essentially an \textit{implicit extrapolation} from the sampled region ($\xi\in(0,1)$) toward the fermionic point ($\xi=-1$).
Therefore, its validity still relies on the same analyticity condition: the continuation path must not be obstructed by LY zeros of $\xi$.
To this end, we analyzed the constant-energy contours in the same model system, using the exact partition function as presented in Sec.~\ref{1Dmod}.
Specifically, we plot $\Delta E$ curves as functions of real $\xi$ and real $T$, where $\Delta E \equiv E_N(T,\xi)-E_N(0,-1)$ represents the energy difference from the referenced fermionic system at 0~K.
In Figs.~\ref{zeross}(a) and (b), we show the results for the $N=4$ and $5$ systems, respectively.
The solid black constant-energy contours in these $\Delta E$-$\xi$-$T$ plots correspond to the trajectories $\xi_{\Delta E}(T)$ in Eq.~\eqref{Xiong}.
The notation using $\Delta E$ instead of $E$ is different from that in Ref.~[\onlinecite{Xiong2023}] but is only a relabeling, since $E_N(0,-1)$ is independent of both $\xi$ and $T$.
In Ref.~[\onlinecite{Xiong2023}], the curves of $\xi_E(T)$ were assumed to follow schematic shapes like insets of Fig.~\ref{zeross}(a)(b).
Our exact results show that the contour geometry depends strongly on $\Delta E$: it resembles the schematic form at high $\Delta E$ but changes qualitatively at low $\Delta E$.
This behavior follows the electrostatic analogy in which LY zeros act as charges: near real-zero trajectories (thick red curves in Fig.~\ref{zeross}), contours bunch up closer together and bend strongly, whereas far from them---either when zeros move off the real axis (thick grey curves in Fig.~\ref{zeross}) or when $\Delta E$ is sufficiently large---the contours become smooth and can be well fitted by Eq.~\eqref{Xiong}.
Therefore, implicit extrapolation based on constant-energy contours and direct extrapolation have the same regime of validity: both fail at low $T$ when the continuation region approaches LY zeros, but remain reasonable at moderate and high $T$.
At low $T$, Fig.~\ref{zeross} directly visualizes this mechanism.
For both $N=4$ and $N=5$, contours near the real-zero trajectories (red curves) are strongly distorted, so data sampled at $\xi\in(0,1)$ cannot be reliably continued to $\xi=-1$ in these regions (e.g., the yellow dashed window for $N=4$).
As further shown in Fig.~\ref{fit}, even a sixth-order expansion based on Eq.~\eqref{Xiong} does not yield a satisfactory estimate of $E_{N=4}(T,\xi=-1)$.
Increasing the expansion order improves agreement at moderate and high $T$, but the low-$T$ error remains large, indicating that high-order fitting still fails in this regime.
Meanwhile, as discussed at the end of Sec.~\ref{1Dmod}, the parity of $N$ constrains the zero trajectories and thus the local contour geometry.
Accordingly, the low-$T$ distortions are parity dependent for $N=4$ versus $N=5$.
The real-zero trajectories form one arc for $N=4$ and two arcs for $N=5$.
In particular, for odd $N$ (e.g., $N=5$), the arc connecting $\xi=-1$ and $-1/2$ distorts a larger region of the contour landscape, as shown in Fig.~\ref{zeross}(b).
By contrast, for even $N$, the trajectory originating from $\xi=-1$ bends toward $\xi>-1$ and induces a weaker distortion.
This even-odd difference weakens with increasing $N$ and becomes negligible in the thermodynamic limit, while the low-$T$ limitation from nearby LY zeros remains.

\begin{figure*}[htbp]
    \centering
    \includegraphics[width=0.95\linewidth]{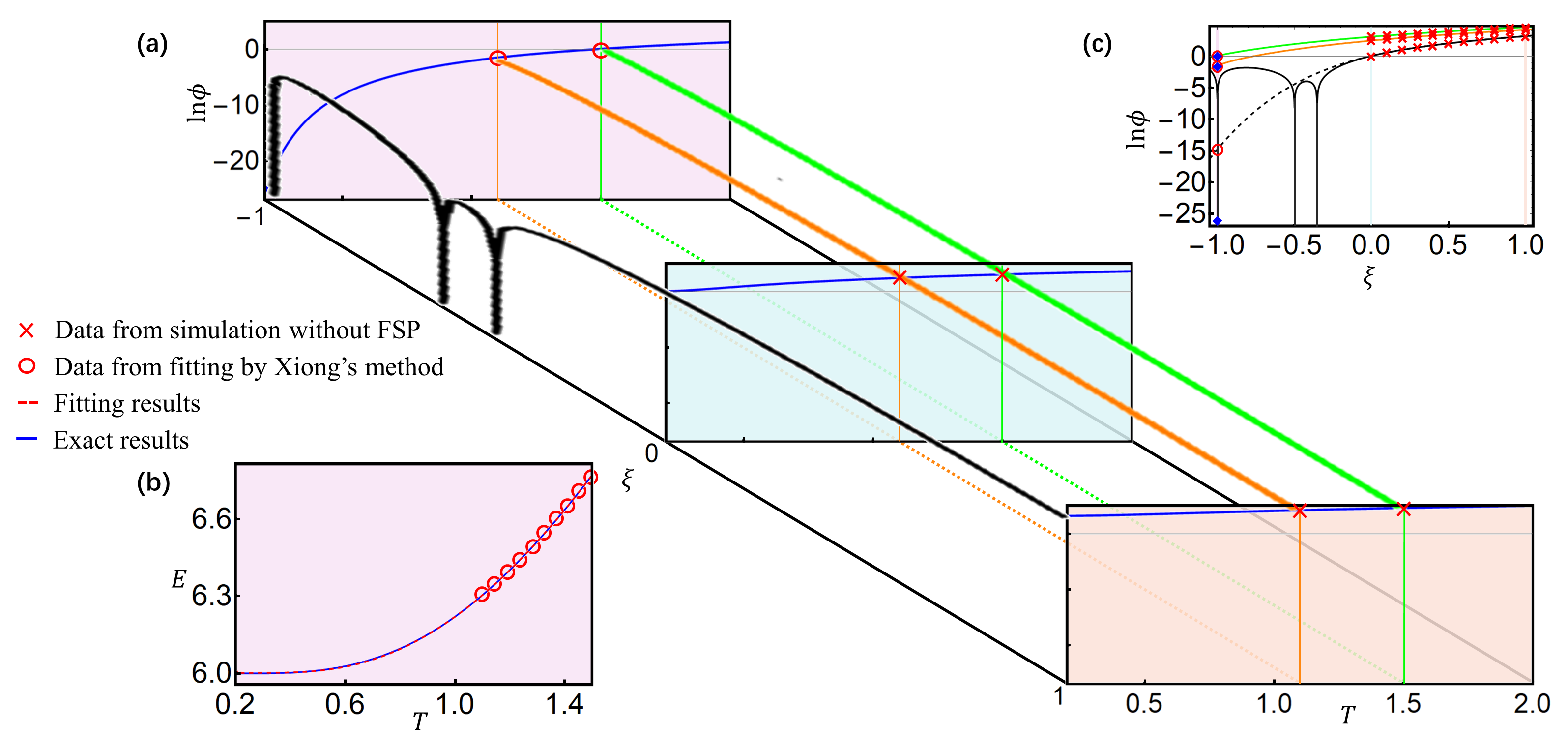}
    \caption{(a) Two-step strategy for accessing low-$T$ fermionic properties. In the first step, sign-problem-free data in $\xi\in[0,1]$ (red crosses) are sampled at high $T$ and extrapolated in $\xi$ to the fermionic point $\xi=-1$; the orange and green lines illustrate such high-$T$ extrapolations, and the hollow circles denote the corresponding estimates at $\xi=-1$. These high-$T$ fermionic estimates are then used for a temperature fit toward lower $T$. (b) The temperature fit (red dashed line), constructed from the high-$T$ fermionic points, agrees closely with the exact energy (blue solid line). (c) The $\xi$ extrapolation fails at $T=1$ (black), as exact result (blue solid diomands) deviates from the fitted one (red hollow circles); but it becomes reasonable at higher $T$, such as $T=1.1$ (orange) and $T=1.5$ (green), where the exact result aligns with the fitted one.}
    \label{fig:tfit}
\end{figure*}

\section{Structure of the partition function}
\label{pfstructure}
The discussion above shows that $z_1$, the LY zero nearest to $\xi=-1$ and exactly equal to $-1$ at $T=0$, strongly affects both the sign factor and the reliability of $\xi$ extrapolation.
This observation motivates us to look more directly at the structure of the partition function around the fermionic point.
For any finite-$N$ system whose partition function can be written as Eq.~\eqref{anyonpoly}, one may divide the polynomial by $\xi+1$ and write
\begin{equation}
    Z(N,\beta,\xi) =(\xi-z_1)\cdot \frac{Z}{\xi-z_1} = (\xi + 1) Z'(N-1, \beta, \xi) + \phi(\beta),
    \label{pfdecomp}
\end{equation}
where $Z' = \sum_{j=0}^{N-2} c_j'\cdot \xi^j$ is an auxiliary polynomial and $\phi(\beta)$ is the $\xi$-independent remainder.
$Z'$ resembles a new partition function with $N-1$ particles, and its zeros are not close to $\xi = -1$.
Intuitively, it may be regarded as an approximation to $Z/(\xi-z_1)$ once $z_1$ is sufficiently close to $\xi = -1$.
The coefficients of $Z$ and $Z'$ are connected by
\begin{equation}
    \begin{aligned}
        c_j &= c_j' + c_{j-1}' && (j=1,\cdots,N-2)\\
        c_{N-1} & =c'_{N-2},\quad c_{0} = c_{0}' + \phi,
    \end{aligned}
\end{equation}
and, conversely,
\begin{equation}    
    \begin{aligned}
    c_{j}' &= \sum_{k=j}^{N-2} (-1)^{k-j}c_{k+1}&& (j=0,\cdots,N-2)\\
    \phi &= c_0 + \sum_{k=1}^{N-1} (-1)^k c_{k} = \sum_{k=0}^{N-1} (-1)^k c_k.
    \end{aligned}
\end{equation}

This decomposition has a simple physical meaning.
Evaluating at $\xi=\pm 1$ gives the fermionic and bosonic partition functions, as
\begin{equation}
    \begin{aligned}
        Z_{\text{F}} &= Z(\xi=-1) = \phi(\beta),\\
        Z_{\text{B}} &= Z(\xi=+1) = 2Z'(N-1,\beta,1) + \phi(\beta).
    \end{aligned}
\end{equation}
Thus $\phi$ is precisely the fermionic partition function $Z_{\text{F}}$, whereas $Z'$ provides the remaining contribution needed to recover the bosonic partition function $Z_{\text{B}}$.
At $0$~K, $\phi=0$ because $z_1=-1$ exactly; at low $T$, $\phi$ remains small because $z_1$ stays close to $-1$.
As implied by Eq.~\eqref{z1}, $z_1 = -1 + \mathcal{O}(e^{-\beta E_0})$, so $\phi$ is expected to be exponentially small at low $T$s.
In an extrapolation scheme, one essentially tries to extract $\phi$ from the sampled region $\xi\in[0,1]$, where $Z$ contains both the desired $\phi$ and undesired contributions from $Z'$.
Especially at low $T$s, the noise from $Z'$ can be significant, overwhelming the desired signal $\phi$ and rendering the extrapolation unreliable.
This gives another view of the FSP: the difficulty is not only failed extrapolation, but also the fact that the fermionic contribution itself becomes exponentially small and therefore hard to sample accurately.
\section{Possible approaches for Fermionic systems}
\label{attempt}
Analysis of the previous sections leads to the conclusion that both direct extrapolation (e.g., Refs.~\cite{Xiong2022,Xiong2023}) and implicit extrapolation (e.g., constant-energy-contour methods) become unreliable at low $T$s.
Nevertheless, this does not mean that $\xi$ extrapolation is useless in all regimes.
Once the relevant LY zeros move away from the real axis at moderate and high $T$s, the continuation from the sign-problem-free region $\xi\in[0,1]$ to the fermionic point $\xi=-1$ becomes reliable again.
The result of Sec.~\ref{pfstructure} then suggests a two-step route for low-$T$ fermionic properties, as summarized in Fig.~\ref{fig:tfit}.
In the first step, high-$T$ data at $\xi\in[0,1]$ are used to obtain $Z_{\mathrm F}=\phi$ or derived fermionic quantities at $\xi=-1$.
In the second step, these high-$T$ fermionic results are fitted as functions of $T$ and continued toward the lower-$T$ regime where direct $\xi$ extrapolation is obstructed.

\begin{figure}[htbp]
    \centering
    \includegraphics[width=\linewidth]{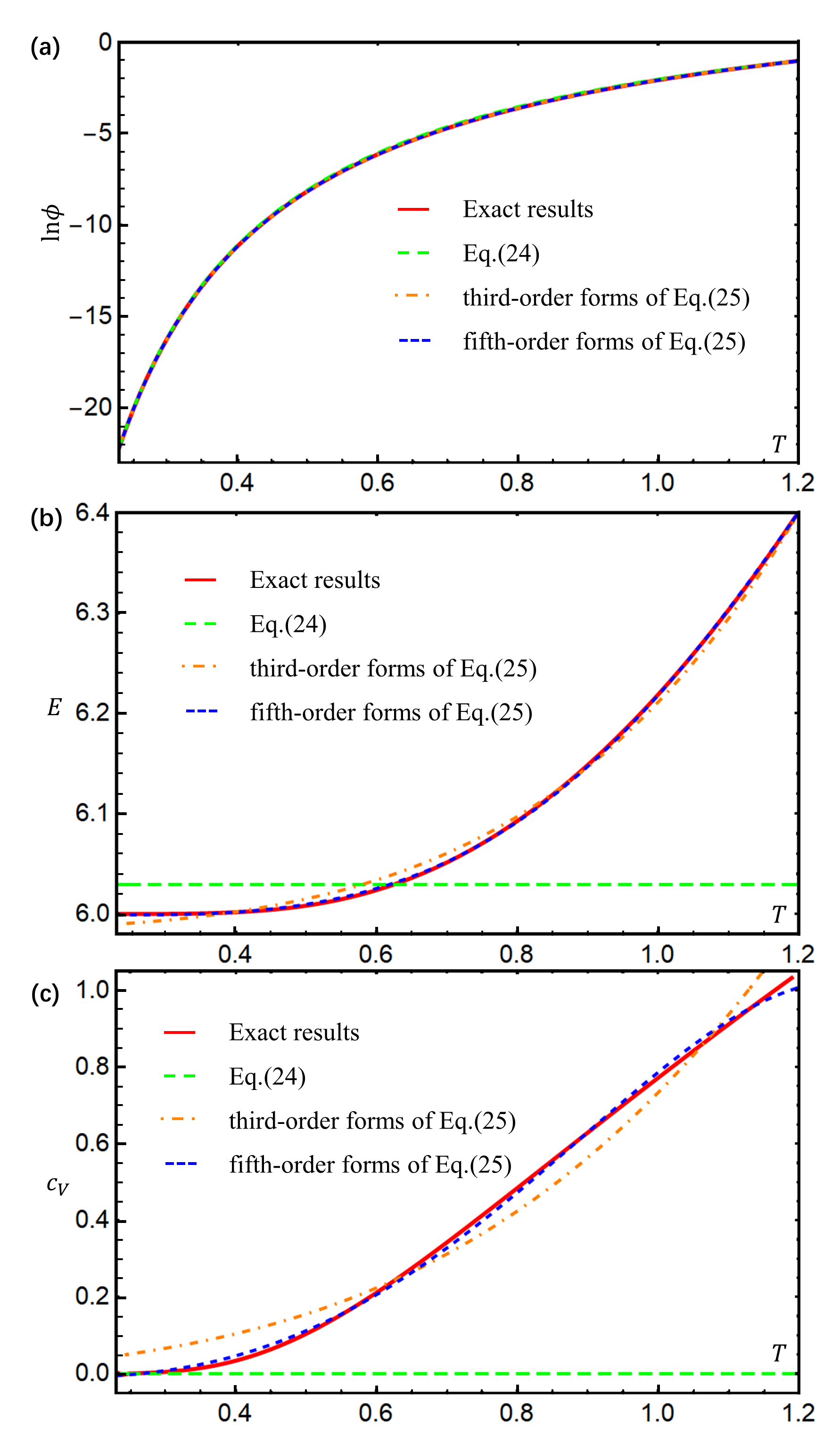}
    \caption{Temperature dependence of (a) $\phi(\beta)=Z_{\mathrm F}$, (b) energy $E$, and (c) heat capacity $c_V$ for the 1D particle-on-a-ring model. Red solid lines show exact results. The green, orange, and blue dashed lines are fitted according to the two-parameter form of Eq.~\eqref{two-param}, the third-order and fifth-order forms of Eq.~\eqref{high-order}, respectively.}
    \label{fig:lnphi}
\end{figure}

More specifically, in the first step, the interval $\xi\in[0,1]$ is free of the sign problem and can therefore be sampled accurately, giving the simulation data marked by red crosses in Fig.~\ref{fig:tfit}(a) and (c).
Using such high-$T$ data, for example those on the orange line at $T=1.1$ and the green line at $T=1.5$, one can extrapolate in $\xi$ and obtain the hollow-circle estimates at $\xi=-1$.
In the second step, these high-$T$ fermionic points are extrapolated toward lower $T$ using the temperature fitting form in Eq.~(\ref{high-order}).
As shown in Fig.~\ref{fig:tfit}(b), the fitted result (red dashed line) using only the high-$T$ data is almost coincident with the exact result (blue solid line) even at low $T$s.
For the present model, Fig.~\ref{fig:lnphi}(a) shows that $\phi(\beta)$ decreases by many orders of magnitude as $T$ is lowered.
Moreover, its logarithm is well described by a simple two-parameter form,
\begin{equation}
    \ln \phi(\beta)=a_1\beta+a_2,
    \label{two-param}
\end{equation}
where $a_1$ is approximately the negative average energy $-E$ over this temperature range.
If one wishes to use the fitted function to study the variation of energy $E=-\partial_\beta\ln\phi$, or even further, the variation of heat capacity $c_V$, higher-order fitting 
\begin{equation}
    \ln \phi(\beta)=a_1\beta+a_2+a_3T+a_4T^3+a_5T^5+\cdots,
    \label{high-order}
\end{equation}
is required to obtain accurate derivative information, as shown in Figs.~\ref{fig:lnphi}(b) and~\ref{fig:lnphi}(c).
Once $\phi$ is reliably reconstructed, accurate fermionic thermodynamic properties can be obtained.

The applicability of this temperature fitting depends on the intrinsic finite-$T$ behavior of the fermionic system.
If the fermionic system undergoes a phase transition, a single smooth fitting function cannot be expected to continue through that nonanalytic point.
If the system instead exhibits only a crossover, or has no phase transition in the temperature window of interest, the temperature fitting of $\phi(\beta)=Z_{\mathrm F}$ remains a viable route, provided that the high-$T$ input data are obtained in a regime where $\xi$ continuation is reliable and the fitting window is chosen appropriately.
This provides a direct route to fermionic properties, although its general applicability beyond the present solvable model remains to be explored.

\section{conclusion}
\label{Conclusion}
In conclusion, we systematically analyzed the finite-$T$ evolution of the Lee-Yang zeros of $\xi$ in an exactly solvable noninteracting model on a one-dimensional ring.
We find that low-$T$ behavior is dominated by the zero near $\xi=-1$, whereas at moderate and high $T$ the relevant zeros move away from the real axis.
This zero evolution explains why both direct and implicit extrapolation on the real $\xi$ axis fail at low $T$, even at high fitting order, but become reasonable again at higher $T$.
Analyzing the polynomial structure of the partition function further shows that the $\xi$-independent remainder after division by $\xi+1$ is exactly the fermionic partition function, $\phi(\beta)=Z_{\mathrm F}$.
This observation leads to a practical low-$T$ strategy: first obtain reliable high-$T$ fermionic information by continuing sign-problem-free $\xi\in[0,1]$ data to $\xi=-1$, and then fit $\phi(\beta)$ or its derived thermodynamic quantities along the temperature direction.
Looking ahead, we expect this strategy to offer a useful route for realistic interacting systems as well.

\begin{acknowledgments}
    We thank Y.N. Xiong and H.W. Xiong for helpful discussions. We are supported by Science Challenge Project, No.TZ2025013, the National Science Foundation of China under Grant No.12550005, 12234001, 12522410, 12474215, and 62321004, andthe National Basic Research Programs of China under Grand Nos. 2021YFA1400500 and 2022YFA1403500. We thank the supercomputer center at Peking University for computational resources. 
\end{acknowledgments}
\bibliography{ref}
\end{document}